\documentclass{aastex631}

\usepackage{graphicx}	
\usepackage{amsmath}	
\usepackage{amssymb}	
\usepackage{float}
\usepackage{graphicx}
\usepackage{subfigure}
\usepackage{multirow}
\def\degree{${}^{\circ}$}

\submitjournal{ApJ}

\shorttitle{TEMPORAL AND SPATIAL BEHAVIORS OF CME OCCURRENCE RATE}
\shortauthors{Lin et al.}
\graphicspath{{./}{figures/}}

\begin{document}

\title{The Temporal and Spatial Behaviors of CME Occurrence Rate at Different Latitudes}

\correspondingauthor{Feng Wang, Linhua Deng, Ying Mei and Hui Deng}
\email{fengwang@gzhu.edu.cn,lhdeng@ynao.ac.cn, meiying@gzhu.edu.cn, denghui@gzhu.edu.cn}

\author{Jiaqi Lin}
\affiliation{Center For Astrophysics, Guangzhou University,
Guangzhou,Guangdong, 510006, P.R. China}
\affiliation{Great Bay Center, National Astronomical Data Center,  
Guangzhou, Guangdong, 510006, P.R. China}

\author{Feng Wang}
\affiliation{Center For Astrophysics, Guangzhou University, 
Guangzhou,Guangdong, 510006, P.R. China}
\affiliation{Great Bay Center, National Astronomical Data Center,
Guangzhou, Guangdong, 510006, P.R. China}
\affiliation{Yunnan Observatories, Chinese Academy of Sciences,
Kunming, Yunnan, 650216, P.R. China}

\author{Linhua Deng}
\affiliation{Yunnan Observatories, Chinese Academy of Sciences,
Kunming, Yunnan, 650216, P.R. China}


\author{Hui Deng}
\affiliation{Center For Astrophysics, Guangzhou University,
Guangzhou,Guangdong, 510006, P.R. China}
\affiliation{Great Bay Center, National Astronomical Data Center,
Guangzhou, Guangdong, 510006, P.R. China}

\author{Ying Mei}
\affiliation{Center For Astrophysics, Guangzhou University, 
Guangzhou,Guangdong, 510006, P.R. China}
\affiliation{Great Bay Center, National Astronomical Data Center,
Guangzhou, Guangdong, 510006, P.R. China}

\author{Yangfan Xie}
\affiliation{Center For Astrophysics, Guangzhou University, 
Guangzhou,Guangdong, 510006, P.R. China}
\affiliation{Great Bay Center, National Astronomical Data Center,
Guangzhou, Guangdong, 510006, P.R. China}



\begin{abstract}


The statistical study of the Coronal Mass Ejections (CMEs) is a hot topic in solar physics. To further reveal the temporal and spatial behaviors of the CMEs at different latitudes and heights, we analyzed the correlation and phase relationships between the occurrence rate of CMEs, the Coronal Brightness Index (CBI), and the 10.7-cm solar radio flux (F10.7). We found that the occurrence rate of the CMEs correlates with CBI relatively stronger at high latitudes ($ \geq 60 $\degree) than low latitudes ($ \leq 50 $\degree). At low latitudes, the occurrence rate of the CMEs correlates relatively weaker with CBI than F10.7. There is a relatively stronger correlation relationship between CMEs, F10.7, and CBI during Solar Cycle 24 (SC24) than Solar Cycle 23 (SC23). During SC23, the high-latitude CME occurrence rate lags behind F10.7 by three months, and during SC24, the low-latitude CME occurrence rate leads the low-latitude CBI by one month. The correlation coefficient values turn out to be larger when the very faint CMEs are removed from the samples of the CDAW catalog. Based on our results, we may speculate that the source regions of the high/low-latitude CMEs may vary in height, and the process of magnetic energy accumulation and dissipation is from the lower to the upper atmosphere of the Sun. The temporal offsets between different indicators could help us better understand the physical processes responsible for the solar-terrestrial interactions.



\end{abstract}



\section{Introduction} 
\label{1}

Coronal Mass Ejections (CMEs), the large-scale bursts of plasma and magnetic fields in the Sun, are the primary sources of catastrophic space weather, which causes many moderate to intense geomagnetic storms \citep{1990GMSGol,2000JGRWebb,gopalswamy2000interplanetary,wang2003multiple}. 

Since the successful launch of the Solar and Heliospheric Observatory (SOHO)~\citep{RN12} as an onboarding instrument, the Large Angle Spectrometric Coronagraph (LASCO) has observed more than 30,000 samples of CMEs. Based on these data, scientists have extensively studied various physical properties of CMEs, such as their speed \citep{RN46, RN47, RN21}, acceleration \citep{RN42, RN47}, occurrence \citep{RN43, RN18}, latitude \citep{RN44, RN47}, mass \citep{RN47, RN18}, and so on. These studies have significantly advanced our understanding about CMEs. 

The statistical study of the temporal and spatial behaviors of CMEs helps to understand the characteristics, formation process, and origin of CMEs, which is essential for us to understand solar activities and predict catastrophic space weather. \citet{RN13} found that the occurrence rate of CMEs tends to track the solar cycle in both amplitude and phase by analyzing the variation of CME rate over time from 1973 to 1989. \citet{RN14} analyzed the relationship between the occurrence rate of CMEs and the sunspot number from 1996 to 2002. They found that the occurrence rate of CMEs is highly correlated with the sunspot number (the coefficient value is 0.86). At the same time, for Solar Cycle 23 (SC23), their peaks are not at the same position but are two years apart. Furthermore, \citet{RN44} showed that a fraction of CMEs originates in spot-free areas.
\citet{RN18} analyzed the difference of the CME occurrence rate between the northern and southern hemisphere and confirmed that there exists asymmetry between the two hemispheres. \citet{2015ApJ...804L..23G, RN82, 2017SoPh..292....5C,2020ApJ...889....1B} found that the number of CMEs during Solar Cycle 24 (SC24) is larger than SC23. \citet{2020ApJ...889....1B} noted that the high-latitude CMEs contributes more to the increase of the number of CMEs during SC24. 

Several studies have found some differences between the high-latitude and low-latitude CMEs. The high-latitude CMEs are closely related to polar crown filaments, while the low-latitude CMEs mainly to sunspots \citep{RN14, Gopalswamy2010}. Moreover, there exist differences in statistical properties between them. \citet{RN21} found that during SC23, the speed of the low-latitude CMEs is slightly larger than the high-latitude CMEs. \citet{2015Gopalswamy} found that during SC24, the low-latitude CMEs own larger speeds, wider angular widths, and smaller accelerations than the high-latitude CMEs. \citet{RN21, Gopalswamy2009, 2015Gopalswamy} concluded that the main factor that causes the CMEs have a larger speed is the magnetic field. Compared to the high-latitude CMEs, the source regions of low-latitude CMEs have higher magnetic-field strength and thus provide higher free energy to power the CMEs. Furthermore, \citet{2011ApJKilcik,2017Kilcik} pointed out that the Maximum CME Speed Index (MCMESI) is a useful indicator of geomagnetic activities. The MCMESI represents CMEs with a larger speed. Statistically, CMEs with larger speeds are often observed at low latitudes \citep{RN21,2015Gopalswamy}. It indicates that the low-latitude CMEs may be a better indicator of geomagnetic activities than the high-latitude CMEs. Therefore, studying the temporal and spatial behaviors of CMEs at different latitudes is instructive for understanding their differences.

Usually, the temporal and spatial behaviors of the CMEs can be studied by comparing the relationship between the CME occurrence rate and solar activity indicators \citep{RN18}. It is well known that the occurrence rate of CMEs varies with the solar cycle. The study of the relationship between CMEs and different solar activity indicators is helpful for connecting the variation of CMEs with the variation of solar activities, which would provide some enlightenment to the source region of CMEs. \citet{RN15} pointed out that the 10.7-cm solar radio flux (F10.7) is a more reliable indicator of solar activities than the number of sunspots because the number of sunspots is determined by observers. F10.7 is directly related to the total amount of magnetic flux that is supposed to come from the lower solar corona which is about 60,000 km above the surface-atmosphere \citep{RN78, RN38, RN79}. Therefore, F10.7 can be considered a low coronal activity index \citep{1990Tapp, RN38}. CBI \citep{RN19}, a new white-light coronal brightness index,  constructed from the LASCO C2 coronagraphs from 1996 to 2017, can be used to explore the spatial relationships between coronal structure and other geophysical indices. CBI contains coronal brightnesses from 2.5 $R_{\odot}$ to 6.2 $R_{\odot}$ coronal heights, which can represent the high coronal activity index. 

We analyzed the correlation and phase relationships between the occurrence rate of CMEs, CBI, and F10.7 to study the temporal and spatial behaviors of CME occurrence rate at high/low latitudes. In the rest of this paper, we introduce data preparation in Section \ref{2}. Section \ref{3} compares the occurrence rate of the low/high-latitude CME occurrence rate with F10.7 and CBI, studies the correlation relationship and phase relationship between CMEs and F10.7, CMEs and CBI, respectively. Section \ref{4} presents conclusions and discussions.

\section{DATA SOURCE AND PRE-PROCESSING} \label{2}

\def\degree{${}^{\circ}$}

\subsection{CME Data Preparation}
\label{2.1}

There are a total of five catalogs of CMEs from SOHO/LASCO \citep{RN12} observations, i.e., CDAW\footnote{https://cdaw.gsfc.nasa.gov/CME\_list/index.html} \citep{RN72,RN73}, ARTEMIS \citep{RN77}, CACTus \citep{RN74,RN75}, SEEDS \citep{RN76} and CORIMP \citep{RN81}. 
CDAW is manually maintained and verified, whereas the other four are automatically generated. Therefore, most statistical analyses on CMEs used the CDAW catalog. We also used the CDAW catalog for our research work. There are 31310 CMEs in the catalog we used, covering from January 1996 to August 2021. CDAW records the physical characteristics of each CME, including time, date, Central Position Angle (CPA), angular width, the 2nd-order speed at final height, accelerated speed, etc..

To study the CMEs at high/low latitudes respectively, we converted CPA to projected heliographic latitude (i.e., the apparent latitude) \citep{RN22, RN20, RN21, RN23, RN24}. For example, CPAs of 0\degree, 90\degree, 180\degree, and 270\degree correspond to the apparent latitudes of 90\degree, 0\degree, $-90$\degree, and 0\degree, respectively. According to the general division to high/low latitudes from the previous study \citep{1998Sakurai,RN24}, we grouped the CMEs whose apparent latitude $ \leq 50 $\degree ~into low-latitude CMEs, whose apparent latitude $ \geq 60 $\degree ~into high-latitude CMEs, and did not research the CMEs whose latitudes are between 50 and 60, considering the projection effect and the magnetic field mixing effect. Since the apparent latitudes of the Halo CMEs (the angular width is 360\degree) can not be determined, we did not use the Halo CMEs (the amount is 734, account for 2.34\% of the total) for our work.

\begin{figure}
\begin{center}
	\includegraphics[width=\columnwidth]{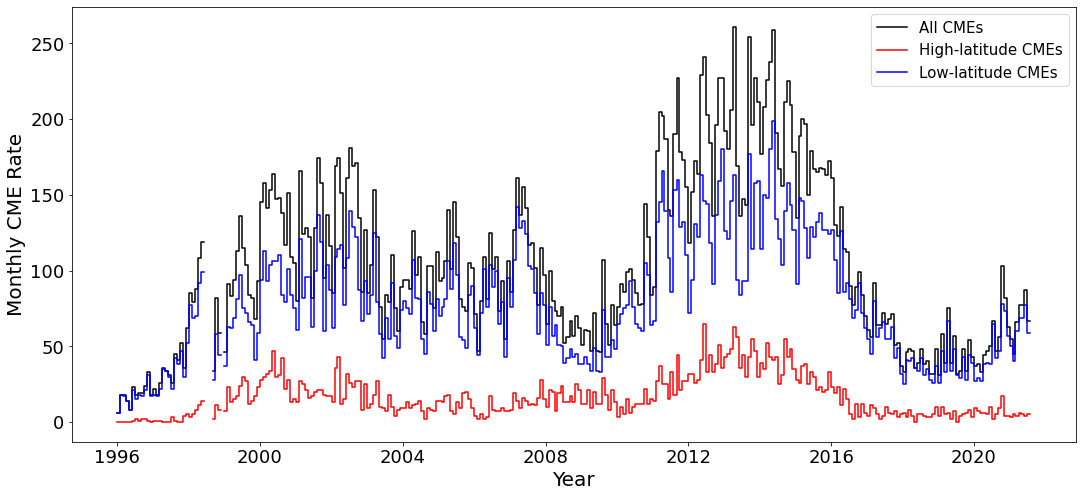}
\end{center}
    \caption{The monthly occurrence rate of CMEs derived from the CDAW catalog. The black line represents all CMEs, the red line represents the high-latitude CMEs, and the blue line represents the low-latitude CMEs.}
    \label{fig:CME_data}
\end{figure}

We finally obtained 4964 high-latitude CMEs (account for 15.9\%) and 23789 low-latitude CMEs (account for 76.0\%). Figure \ref{fig:CME_data} shows the time series of the monthly occurrence rate of CMEs for all/high/low latitudes. As can be seen from Figure \ref{fig:CME_data}, compared with the high-latitude CMEs, the low-latitude CMEs account for the vast majority.

\subsection{CBI Data Preparation}
\label{2.2}
\begin{figure}
\begin{center}
	\includegraphics[width=\columnwidth]{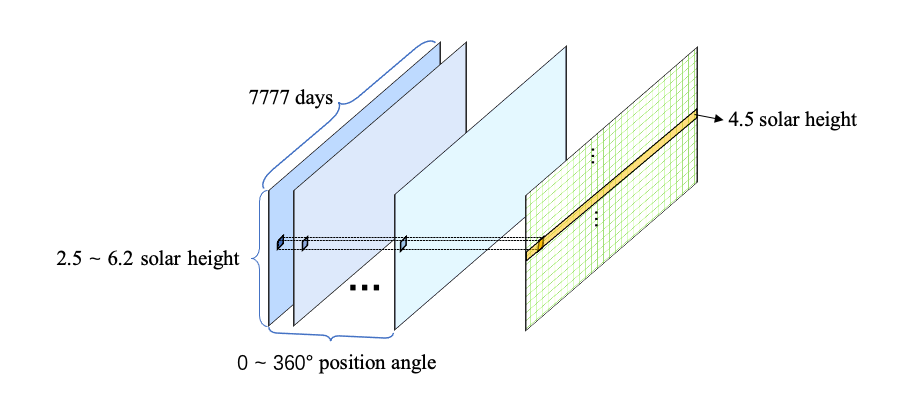}
\end{center}
    \caption{Schematic diagram of the CBI data. The yellow-green part on the right are the accumulations along the position angle. The line in yellow color represents the data at the  height of 42.5 $R_{\odot}$.}
    \label{fig:CBI_type}
\end{figure}

The CBI data is obtained from the website \footnote{https://lasco-www.nrl.navy. mil/CBI} generated by \citet{RN19}. The time is from 16 April 1996 to 31 July 2017, including 7777 days. There are two datasets of CBI: the raw dataset (the data is not continuous because there are some days that have no CBI values) and the interpolated dataset (those days without values have been assigned values through linear interpolation). We used the interpolated dataset for our research. The data cube is 360 $\times$ 38 $\times$ 7777 (the blue grey part in Figure \ref{fig:CBI_type}). The first dimension (i.e., 360) is the position angle (along the limb of the Sun), which starts at a position angle of 0\degree ~(corresponding to solar north), then passes counterclockwise through the LASCO field of view so that 90\degree ~corresponds to the eastern limb, 180\degree ~to the solar south, 270\degree ~to the western limb, and goes back to 0\degree~ in the end. The second dimension (i.e., 38) represents the height in the corona that is from 2.5 $R_{\odot}$ to 6.2 $R_{\odot}$ with a step of 0.1 $R_{\odot}$. The third dimension (i.e., 7777) represents the days from 16 April 1996 to 31 July 2017, with a step of one day. 

\begin{figure}
\begin{center}
	\includegraphics[width=\columnwidth]{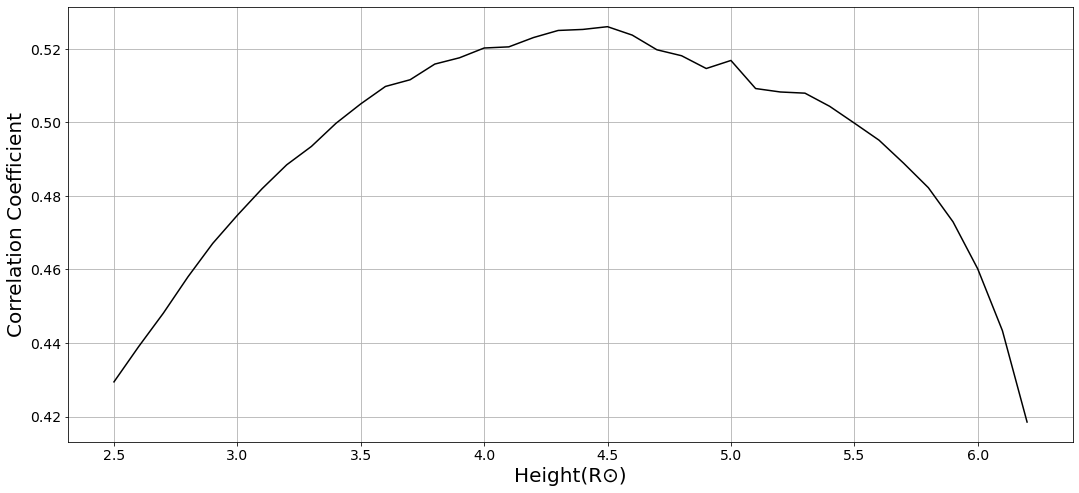}
\end{center}
    \caption{The correlation coefficients between the monthly CME occurrence rate and CBI. The horizontal axis represents the heights that are from 2.5 $R_{\odot}$ to 6.2 $R_{\odot}$ and the vertical axis represents the correlation coefficient values.}
    \label{fig:CME_CBI_CC}
\end{figure}

To find out a proper height of the CBI for our research, we first summed the CBI values along the position angle to obtain a time series of CBI at different heights, i.e., a two-dimensional array (38 x 7777, the yellow-green part in Figure \ref{fig:CBI_type}), then calculated the correlation coefficients between the CME occurrence rate (with all-latitude CMEs) and CBI from different heights to find out the height with the maximum correlation coefficient, i.e., the 4.5 $R_{\odot}$ height. The result is shown in Figure \ref{fig:CME_CBI_CC}. We used the CBI data at the 4.5 $R_{\odot}$ height for our work.


\begin{figure}
\begin{center}
	\includegraphics[width=\columnwidth]{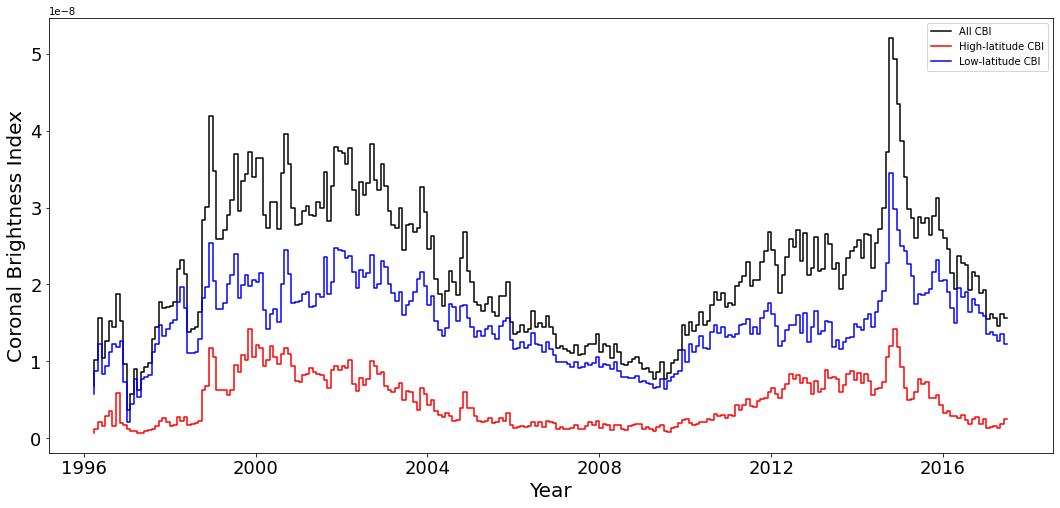}
\end{center}
    \caption{The monthly averaged CBI index. The period is from April 1996 to July 2017.}
    \label{fig:CBI_data}
\end{figure}

Like the CMEs, the CBI also need to be divided into three groups, i.e., all CBI, high-latitude ($ \geq 60 $\degree) CBI, and low-latitude ($ \leq 50 $\degree) CBI. Because the CBI data from the website provides position angle rather than latitude, we should also transform position angles into latitudes with the same way used for CMEs. Figure \ref{fig:CBI_data} shows the time series of monthly averaged CBI index for all latitude, high latitude, and low latitude.

\subsection{Data Source of F10.7}
\label{2.3}
\begin{figure}
\begin{center}
	\includegraphics[width=\columnwidth]{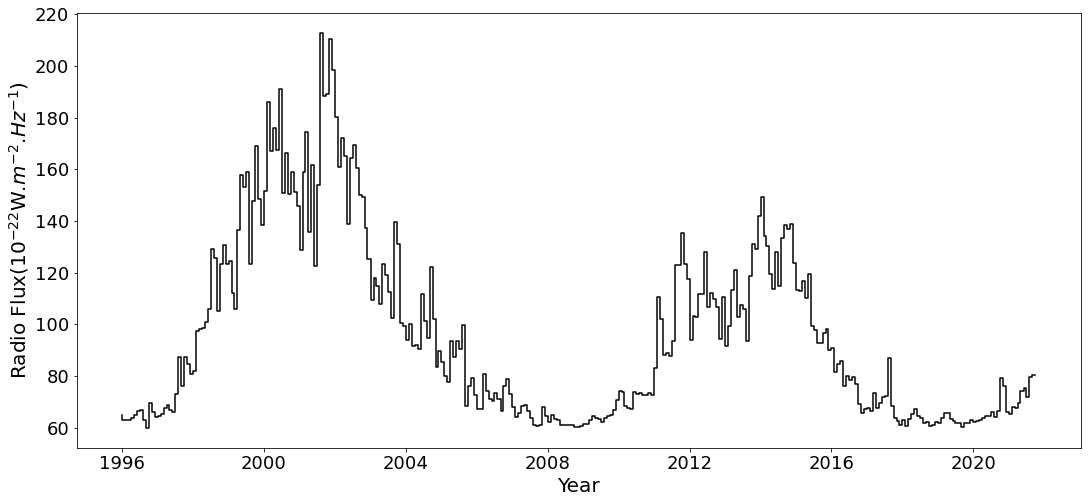}
\end{center}
    \caption{Time series of monthly average values of F10.7 from January 1996 to August 2021.}
    \label{fig:F107_data}
\end{figure}

F10.7, one of the most widely used indicators of solar activity \citep{RN33}, is the measurement of the integrated emission at 10.7cm from all the sources on the disc. Compared to the CBI, F10.7 reflects more of the lower solar corona (about 60,000 km above the solar surface). The data of F10.7 used in our work comes from the Canadian Solar Radio Monitoring Program (Ottawa) \footnote{https://www.spaceweather.gc.ca/forecast-prevision/solar-solaire/solarflux/sx-5-en.php} that gives three types of F10.7 flux: the observed, the adjusted, and the absolute. The absolute F10.7 flux eliminates the fluctuation caused by the variation of Sun-Earth distance and suppresses the antenna gain and ground reflected radiation. We used the absolute flux of F10.7 for our research which covers from January 1996 to August 2021 (see the monthly average F10.7 in figure \ref{fig:F107_data}). The unit of F10.7 is the solar flux unit (sfu), where 1 sfu = $10^{-22} W m^2 Hz^{-1}$.

\section{ANALYSES AND RESULTS} \label{3}

\subsection{Correlation analysis between CME occurrence rate, F10.7, and CBI}
\label{3.1}
\begin{figure}
\begin{center}
	\includegraphics[width=\columnwidth]{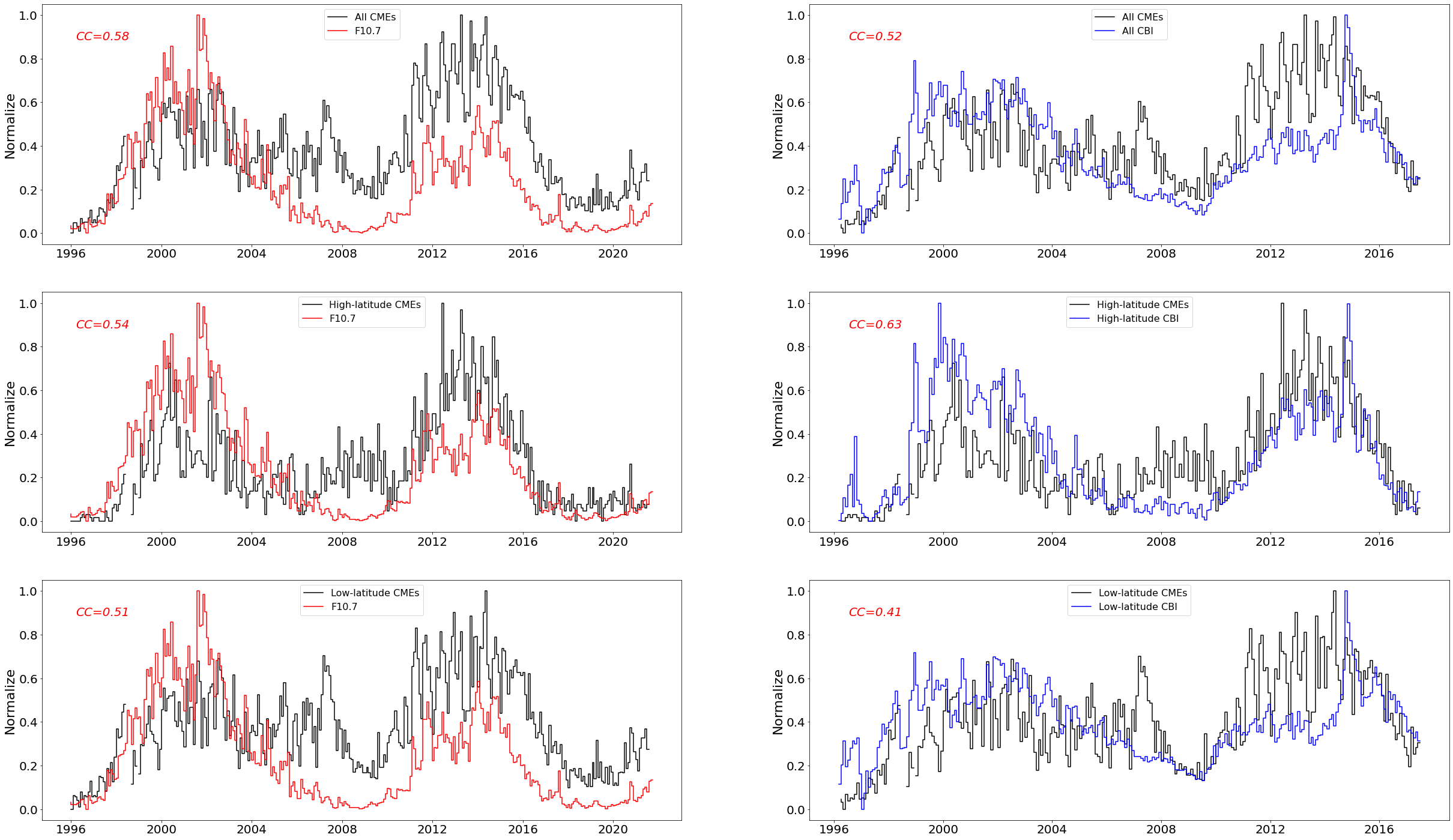}
\end{center}
    \caption{The temporal variations of the monthly CME occurrence rate, the monthly averaged F10.7, and the monthly averaged CBI. The left panels are for the CMEs and F10.7. The right panels are for the CEMs and CBI. From top to bottom are all, high latitude, low latitude, respectively. The red-colored "CC" is the value of the correlation coefficient.}
    \label{fig:CC}
\end{figure}

We explored the correlation relationship between the monthly CME occurrence rate and monthly averaged F10.7 from January 1996 to August 2021, the correlation relationship between the monthly CME occurrence rate and monthly averaged CBI from April 1996 to July 2017. The results are shown in Figure \ref{fig:CC} and Table \ref{tab:1}. The monthly CME occurrence rate, the monthly F10.7, and the monthly CBI are all normalized. In Figure \ref{fig:CC}, the "CC" in red color represents the Correlation Coefficient (CC). The left panels are for the CMEs and F10.7, and the right panels are for the CMEs and CBI. 



From Figure \ref{fig:CC} and Table \ref{tab:1}, we can find that the CME occurrence rate is correlated with F10.7 and CBI. Their correlation relationships at high latitudes are stronger than at low latitudes. For the high latitude, the coefficient value of CC between CMEs and CBI is 0.63, whereas for the low latitude, the value is 0.41. The correlation between high-latitude CMEs and CBI (CC = 0.63) is stronger than that between CMEs and F10.7 (CC=0.54). In contrast, at low latitudes, the correlation between CMEs and CBI (CC = 0.41) is weaker than that between CMEs and F10.7 (CC=0.51).

According to previous research, the number of the CMEs during SC24 is larger than SC23 \citep{2015ApJ...804L..23G, RN82, 2017SoPh..292....5C,2020ApJ...889....1B}, the CMEs during SC24 have lower velocities  \citep{RN18,2020ApJ...889....1B} and weaker photospheric magnetic fields \citep{2015ApJ...812...20T}, which suggests that the characteristics of the source regions of CMEs during SC24 may be different from that during SC23 \citep{article1}. Therefore, we calculated the correlation coefficients for SC23 and SC24, respectively. The calculation results are listed in Table \ref{tab:1}. From Table \ref{tab:1}, we can find that their correlation relationship during SC24 is stronger than SC23. It is worth noting that, at low latitudes during SC23, the three indices are all weakly correlated (CC=0.43 and 0.39, respectively).

\begin{table}
\centering
\renewcommand{\arraystretch}{1.0}
\caption{The values of the correlation coefficients between CMEs, F10.7, and CBI, during the whole time, SC23, and SC24, respectively. The values in parentheses are the correlation coefficients without the ''Very Poor'' CMEs.}
\label{tab:1}
\begin{tabular}{ccccc}
\hline
                      &       & All        & High-latitude & Low-latitude \\ \hline
1996.1-2021.8         & F10.7 & 0.58(0.79) & 0.54(0.75)    & 0.51(0.75)   \\
1996.4-2017.7         & CBI   & 0.52(0.70) & 0.63(0.81)    & 0.41(0.59)   \\ \hline
\multirow{2}{*}{SC23} & F10.7 & 0.58(0.84) & 0.60(0.80)    & 0.43(0.77)   \\
                      & CBI   & 0.51(0.78) & 0.65(0.83)    & 0.39(0.71)   \\ \hline
\multirow{2}{*}{SC24} & F10.7 & 0.89(0.91) & 0.82(0.87)    & 0.84(0.87)   \\
                      & CBI   & 0.67(0.63) & 0.80(0.83)    & 0.51(0.46)   \\ \hline
\end{tabular}
\end{table}

\subsection{Phase relationship between CME occurrence rate, F10.7, and CBI}
\label{3.2}


We used the Cross-Correlation Analysis (CCA) \footnote{https://www.statsmodels.org/dev/generated/statsmodels.tsa.stattools.ccf.html} to analyze the phase relationship between the monthly CME occurrence rate, monthly averaged F10.7, and monthly averaged CBI  at high latitudes, low latitudes, and all respectively. In the CCA method, the leading or lagging relationship of two time series is determined by the phase lag position of the Maximum Correlation Coefficient (MCC) \citep{2004Hagino,2017Kilcik,2018Kilcik, Deng_2019, Deng_2020}.

The values of MCC and its corresponding phase lag time are listed in Table \ref{tab:2}. To estimate the error level of these MCCs, we applied the Fisher's test to calculate their confidence intervals at 95\% significance \citep{2011ApJKilcik,2017Kilcik,2018Kilcik,2021Tirnakci,2021Ozguc} and the results are listed in Table \ref{tab:2}.

\begin{figure}
\begin{center}
	\includegraphics[width=\columnwidth]{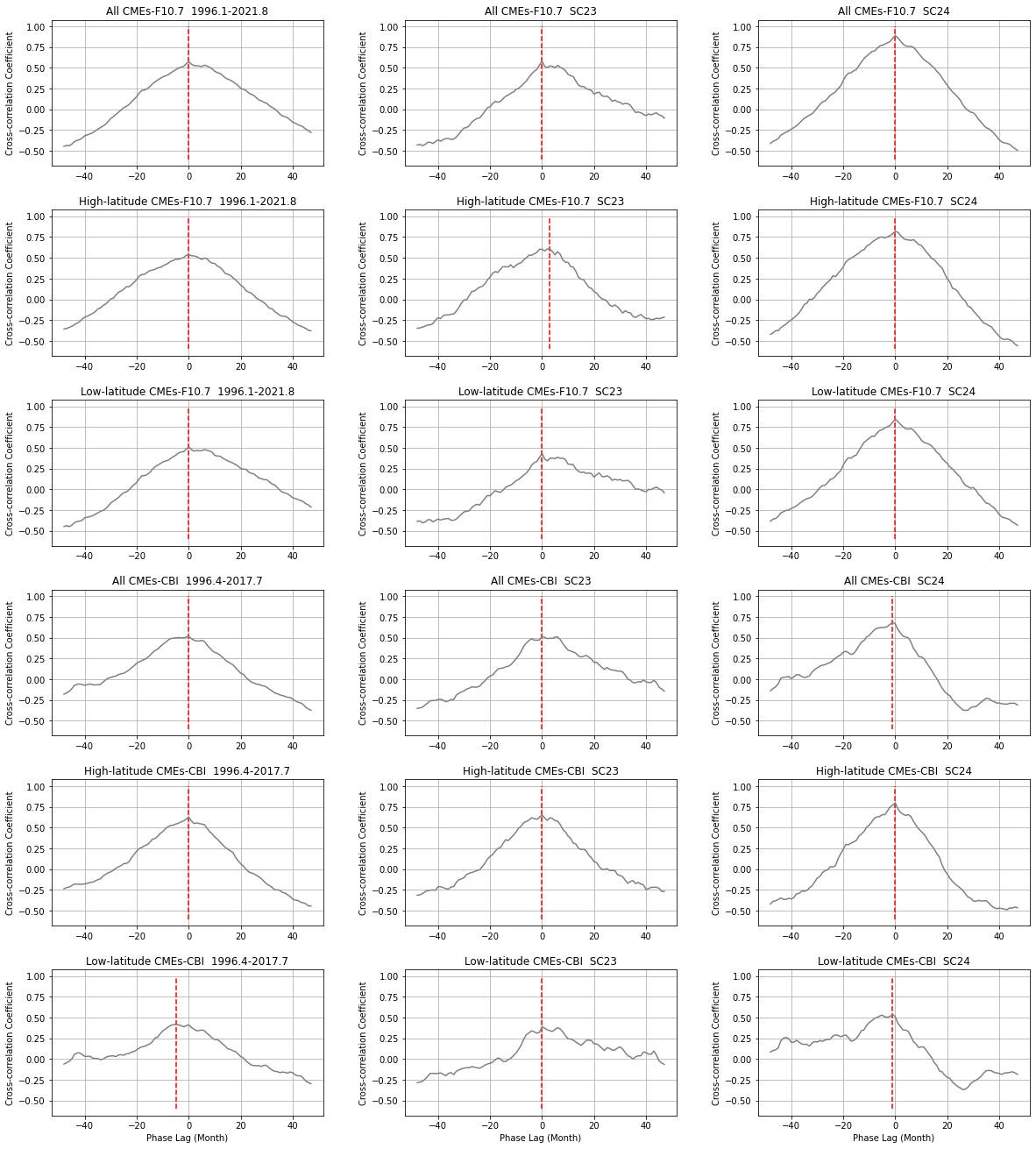}
\end{center}
    \caption{The cross-correlation analysis, i.e., CCA, of the monthly CME occurrence rate, monthly averaged F10.7, and monthly averaged CBI. The left/middle/right panels are for the whole time/SC23/SC24, respectively. The red dotted lines represent the position of the maximum correlation coefficient , i.e., MCC.}
    \label{fig:CCA_cme_rf_cbi}
\end{figure}

From Figure \ref{fig:CCA_cme_rf_cbi} and Table \ref{tab:2}, we can find that in the whole time, the occurrence rate of the high-latitude CMEs is in phase with the high-latitude CBI and F10.7. The occurrence rate of the low-latitude CMEs is in phase with F10.7, whereas leads the low-latitude CBI by five months. During SC23, the occurrence rate of the high-latitude CMEs lags behind F10.7 by three months. During SC24, the occurrence rate of the low-latitude CMEs leads the low-latitude CBI by one month. These results further indicate that the high-latitude CMEs correlate better with the high-latitude CBI. In contrast, the low-latitude CMEs correlate better with F10.7.


\begin{table}
\centering
\renewcommand{\arraystretch}{1.0}
\caption{The values of MCCs with confidence intervals and their corresponding phase lags (unit is month). Negative values indicate that the CME occurrence rate leads the F10.7/CBI, and positive values mean lagging.}
\label{tab:2}
\begin{tabular}{cccccccc}
\hline
                      &            & \multicolumn{2}{c}{All}                    & \multicolumn{2}{c}{High-latitude}            & \multicolumn{2}{c}{Low-latitude}           \\ \hline
                      &            & MCC & Phase Lag(Month) & MCC & Phase Lag(Month) & MCC & Phase Lag(Month) \\ \hline
1996.1-2021.8         & CMEs-F10.7 & 0.58±0.07               & 0                & 0.54±0.08                 & 0                & 0.51±0.08               & 0                \\
1996.4-2017.7         & CMEs-CBI   & 0.52±0.09               & 0                & 0.63±0.08                 & 0                & 0.42±0.10               & -5               \\ \hline
\multirow{2}{*}{SC23} & CMEs-F10.7 & 0.58±0.11               & 0                & 0.61±0.11                 & 3               & 0.43±0.14               & 0                \\
                      & CMEs-CBI   & 0.51±0.13               & 0                & 0.65±0.10                   & 0                & 0.39±0.14               & 0                \\ \hline
\multirow{2}{*}{SC24} & CMEs-F10.7 & 0.89±0.04               & 0                & 0.82±0.06                 & 0                & 0.84±0.05               & 0                \\
                      & CMEs-CBI   & 0.69±0.11               & -1               & 0.80±0.08                 & 0                & 0.54±0.15               & -1               \\ \hline
\end{tabular}
\end{table}

\section{CONCLUSIONS AND DISCUSSIONS} \label{4}

To reveal the temporal and spatial behaviors of the CMEs at different latitudes, we investigated the correlation and phase relationships of their occurrence rate with F10.7 and CBI, and obtained the following results:

\begin{itemize}
\setlength{\itemsep}{0pt}
\setlength{\parsep}{0pt}
\setlength{\parskip}{0pt}

  \item [(1)] 
  The high-latitude CME occurrence rate has a relatively stronger correlation relationship with high-latitude CBI than F10.7 (lower corona);
  
  
  \item [(2)]
    The low-latitude CME occurrence rate has a relatively stronger correlation relationship with F10.7 than low-latitude CBI;
  
  
  
  \item [(3)]
  There is a relatively stronger correlation relationship between CMEs, F10.7, and CBI for SC24 than SC23;
  
  \item [(4)]
For SC23, the high-latitude CME occurrence rate lags behind F10.7 by three months, whereas for SC24, they are in phase. For SC24, the low-latitude CME occurrence rate leads the low-latitude CBI by one month, whereas for SC23, they are in phase.
  
  
\end{itemize}


The occurrence rate of the high-latitude CMEs correlates more strongly with the high-latitude CBI. In contrast, the occurrence rate of the low-latitude CMEs is more correlated with F10.7. F10.7 is a low corona activity index and reflects the global variation of the Sun. CBI is an index that reflects the high corona activities. Therefore, we could speculate that the source regions of high/low-latitude CMEs may vary in height: the source regions of high-latitude CMEs are closer to high coronas, and the source regions of low-latitude CMEs are closer to low coronas.

It is generally accepted that due to the upward evolution of solar activities, the cycles of upper activity indices commonly lag behind the lower activity indices by several months\citep{RN68, Du2011, RN69}. \citet{Bachmann1994} found that six solar activity indicators (F10.7, Ca II K index, equivalent width of He I 1083 nm, Mg II 280 nm core-to-wing ratio, extreme ultraviolet radiation, and L$\alpha$ emission) significantly lag behind the sunspot number by one month to several months during solar cycles 21 and 22. \citet{DENG20121425} found that the sunspot number appears one month earlier than the coronal index. \citet{Tang2013} found that the flare index lags behind the sunspot number and sunspot area. In our work, we find that the CME occurrence rate leads the CBI at low latitudes and lags behind F10.7 at high latitudes. Since CBI reflects the high corona activities and F10.7 reflects the low corona activities, our results further indicate that the process of magnetic energy accumulation and dissipation is from the lower to the upper atmosphere of the Sun.

The existing literature shows that the phase relationships between solar activity indices are different during different solar cycles. \citet{RN71} found there exist significant time lags between flare activities (X-ray flare, H$\alpha$ flare) and sunspot activities (sunspot number, sunspot area) during SC19, SC21, and SC23, whereas they are in phase during SC20 and SC22. \citet{RN70} found that the flare index lags behind the sunspot number during SC21, the flare index is in phase with the sunspot number during SC22, and the flare index leads the sunspot number during SC23. In our work, we find that during SC23, the high-latitude CME occurrence rate lags behind F10.7 by three months, whereas during SC24, the two indices are in phase. During SC23, the CME occurrence rate is in phase with CBI, whereas during SC24, it leads CBI by one month. Our results support that the leading/lagging relationship changes from one cycle to another.

 Based on the ARTEMIS catalog, \citet{RN18} investigated the correlation relationships between CME occurrence rate and F10.7 and other four solar activity indicators (sunspot number, sunspot area, X-ray, total photospheric magnetic flux). They found that their correlation relationships are stronger during SC24 than SC23. Based on the CDAW catalog, we also find that the correlation relationships of CME occurrence rate, F10.7, and CBI are stronger during SC24 than SC23, which agrees with their findings. The difference in the correlation degree during different cycles might be associated with the real differences in the behaviors of strong and weak photospheric magnetic fields in cycles 23 and 24, as pointed out by \cite{2020ApJ...889....1B}.
 
 However, our coefficient values ( 0.58 in SC23, 0.89 in SC24) based on CDAW are smaller than their values (0.93 in SC23, 0.95 in SC24) based on ARTEMIS. We think the reason is that the CDAW catalog is manually maintained, in which some small and faint ejections are included (marked with "Very Poor Event"). In contrast, the ARTEMIS catalog is automatically generated based on a pre-set threshold, in which those small and faint ejections have been dropped. The small and faint ejections in the CDAW catalog are marked since the middle of SC23, which accounts for 28\% of the total CMEs. We also computed the correlation coefficients without these "Very Poor Events." The results are shown in Table \ref{tab:1} (the values in parentheses). We can easily notice that almost all of the correlation coefficients without the "Very Poor Event" increase. The coefficient values of SC23 (0.84) and SC24 (0.91) get close to the results of \citet{RN18}.

\begin{acknowledgments}

This work is supported by the National SKA Program of China No 2020SKA0110300, the Joint Research Fund in Astronomy (U1831204, U1931141) under cooperative agreement between the National Natural Science Foundation of China (NSFC) and the Chinese Academy of Sciences (CAS), National Science Foundation for Young Scholars (11903009). Funds for International Cooperation and Exchange of the National Natural Science Foundation of China (11961141001). Fundamental and Application Research Project of Guangzhou(202102020677). The Innovation Research for the Postgraduates of Guangzhou University (2021GDJC-M13). This work is also supported by Astronomical Big Data Joint Research Center, co-founded by National Astronomical Observatories, Chinese Academy of Sciences and Alibaba Cloud.

We thank the anonymous reviewers for their careful reading of our manuscript and their many insightful comments and suggestions.

\end{acknowledgments}

\bibliography{referrence}



\end{document}